\documentclass[twocolumn,showpacs,prl,floatfix]{revtex4}
\usepackage{epsfig}

\newcommand{\be}{\begin{equation}}
\newcommand{\ee}{\end{equation}}
\newcommand{\beqn}{\begin{eqnarray}}
\newcommand{\eeqn}{\end{eqnarray}}

\begin{document}

\title{Reversal-Field Memory in the Hysteresis of Spin Glasses}
\author{H.~G.~Katzgraber$^1$, F.~P\'azm\'andi$^1$, C.~R.~Pike$^2$,
Kai Liu$^1$, R.~T.~Scalettar$^1$, K.~L.~Verosub$^2$, G.~T.~Zim\'anyi$^1$}
\affiliation{$^1$Department of Physics, University of California, Davis,
California 95616\\ $^2$Department of Geology, University of California, 
Davis,
California 95616}

\date{\today}

\begin{abstract}
We report a novel singularity in the hysteresis of
spin glasses, the reversal-field memory effect, which
creates a non-analyticity in the magnetization curves  
at a particular point related to the history of the   
sample. The origin of the effect is due to the
existence of a macroscopic number of ``symmetric    
clusters'' of spins associated with a local
spin-reversal symmetry of the Hamiltonian. We use  
First Order Reversal Curve (FORC) diagrams to
characterize the effect and compare to experimental
results on thin magnetic films. We contrast our   
results on spin glasses to random magnets and show
that the FORC technique is an effective ``magnetic
fingerprinting'' tool.
\end{abstract}

\pacs{75.50.Lk, 75.40.Mg, 05.50.+q}
\maketitle

The non-equilibrium behavior of random magnets and spin glasses is an
intensely studied field, posing formidable theoretical and experimental
challenges directly for magnetic systems, and also serving as paradigms
for other fields. Concepts developed for random magnets such as glassy
phases, droplet and replica theories, as well as aging have subsequently
been applied to fields as diverse as structural biology, geology, and even
financial analysis.

The slow and complex time dependence of various correlators is a hallmark
of such systems.  Several aspects of this non-equilibrium dynamics have
already been described in great detail for spin glasses \cite{young:98}.
Hysteresis is one of the most central of these
phenomena \cite{bertotti:98}, yet while many basic features are
qualitatively understood \cite{sethna:93,lyuksyutov:99,zhu:90}, theoretical
descriptions of hysteresis even in the simplest spin-glass models are in
their early stages \cite{pazmandi:99}. Hysteresis in magnetic systems has a
host of practical applications including magnetic recording and sensors,
but a less than complete understanding at a fundamental
level \cite{bertotti:98}.

In this paper, we present a detailed study of several new aspects of
hysteresis in two of the most commonly studied models of disordered
magnets.  One is the Random Field Ising Model (RFIM), which has been shown
to describe successfully many of the relevant aspects of
hysteresis \cite{sethna:93}. The second is the Edwards-Anderson Ising Spin
Glass (EASG), which, unlike the RFIM, contains frustration, a
phenomenon known to introduce a whole new level of complexity in
disordered systems. Accordingly, we show that the hysteretic properties of
the EASG can be significantly different from those of the RFIM.

Our first important observation is of a novel memory effect in the
hysteresis of the EASG that emerges when the magnetic field is first
decreased from its saturation value and then increased again from some
reversal field $H_R$. We find that the EASG exhibits a singularity at the
negative of the reversal field, $-H_R$, in the form of a kink in the
magnetization of the reversal curve. By calculating a suitable overlap
function, we demonstrate that the microscopic origin of the effect is due
to a macroscopic number of ``symmetric clusters''. In these clusters the
central spins flip {\it after} all spins on the cluster surface have
flipped. Therefore, the central spins experience an effective local field
which is symmetric with respect to the change of direction of the external
field.

This reversal-field memory effect can be even more precisely characterized
with the recently introduced First Order Reversal Curve (FORC)
method \cite{pike:99}. As we shall demonstrate, the FORC technique
provides a uniquely sensitive characterization of hysteretic systems and
specifically of the difference between the hysteretic behavior of the RFIM
and the EASG. The sharp kink of the minor loops of the EASG is captured as
a profound horizontal ridge in FORC diagrams, indicative of a broad range
of effective coercivities in the system, but a rather narrow range of
biases. In contrast, despite exhibiting a major hysteresis loop rather
similar to that of the EASG, the RFIM shows a strikingly different FORC
diagram, characterized by a well developed vertical feature reflecting a
rather narrow range of effective coercivities and a broad range of biases.

Finally, we determine experimentally the reversal curves and FORC diagram
of a magnetic thin film.  Experimentally, the reversal curves only show
smoothed kinks around $-H_R$.  However, the FORC diagram of the data
reveals a profound horizontal ridge, signaling the presence of a reversal
field memory in these films. This experimental result further highlights
the usefulness of the FORC technique as a powerful method which captures
the detailed behavior of hysteretic systems.

The Hamiltonian of the EASG is given by \cite{binder:86}
\begin{equation}
{\cal H}= \sum_{\langle i,j \rangle} J_{ij}S_iS_j -
                  H  \sum_i S_i \,.
\label{eq:hamilton}
\end{equation}
Here $S_i = \pm 1$ are Ising spins on a square lattice of size $N = L
\times L$ in two dimensions with periodic boundary conditions. The
exchange couplings $J_{ij}$ are random nearest-neighbor interactions
chosen according to a Gaussian distribution with zero mean and standard
deviation unity, and $H$ is the external magnetic field. We simulate the
zero temperature dynamics of the EASG by changing the external field $H$
in small steps, first downward from positive saturation and then upward
from a reversal field $H_R$. After each field step, the effective local
field $h_i$ of each spin $S_i$ is calculated:
\begin{equation}
h_i=\sum_{j} J_{ij}S_j - H \; .
\label{eq:local_field}
\end{equation}
A spin is unstable if $h_i S_i < 0$. We then flip a randomly chosen
unstable spin and update the local fields at neighboring sites and repeat
this procedure until all spins are stable.

\begin{figure}
\epsfxsize=7.0cm
\begin{center}
\epsfbox{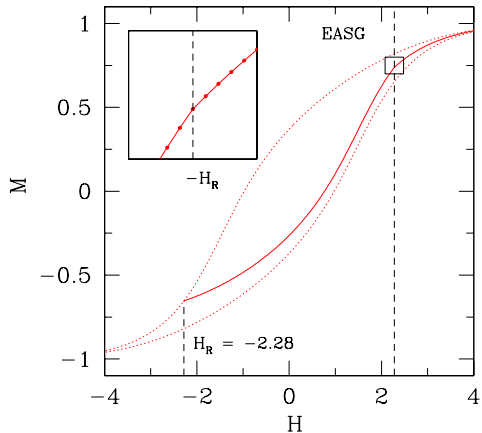}
\end{center}
\vspace*{-1.3cm}
\caption{\label{fig:kink}
Reversal curve (solid line) and major hysteresis loop (dotted line) for a
two-dimensional (2D) EASG with $10^4$ spins and $H_R = -2.28$. In the inset a
kink is seen around $-H_R$. In all figures the error bars are smaller than
the symbols.}
\end{figure}

Figure \ref{fig:kink} (solid line) shows the average of $10^3$ reversal
curves, all with the same $H_R$, but different disorder realizations. The
area around $-H_R$ is enlarged in the inset and shows a ``kink.'' The
presence of any such sharp feature in a disordered system, especially of
finite size and after disorder averaging, is quite remarkable.

\begin{figure}
\epsfxsize=7.0cm
\begin{center}
\epsfbox{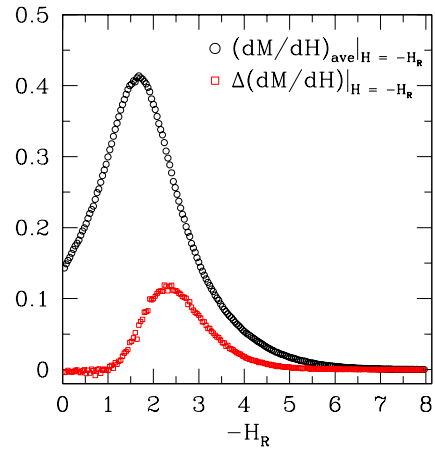}
\end{center}
\vspace*{-1.3cm}
\caption{\label{fig:delta}
The difference (squares) and average (circles) of the left and right
derivatives at $-H_R$ for the EASG.}
\end{figure}

The change of slope at the kink can be characterized by measuring the
slope of the magnetization curves to the left and right of $-H_R$, and
comparing the difference $\Delta (dM/dH)$ with the average $(dM/dH)_{ave}$
(see Fig.~\ref{fig:delta}). The slope changes abruptly by as much as $30
\%$ as the field $H=-H_R$ is passed, creating the kink. With our
parameters the kink is present in the range of reversal-field values $-4.0
< H_{R} < -1.5$.

In an effort to understand the microscopic origin of reversal-field
memory, we first describe this effect within a phenomenological approach
to hysteretic systems, the Preisach model \cite{preisach:35}. In the
Preisach model a magnetic system is described as a collection of
independent two-state ($\pm 1$) switching units, or ``hysterons''. Unlike
Ising spins, which always align with their local field, the hysteron's
state changes from $-1$ to $+1$ at a field $H_b+H_c$, different from the
field $H_b-H_c$, required to switch the hysteron from $+1$ to $-1$.
Different systems are distinguished by their different distributions
$\rho(H_b, H_c)$ of hysterons of a given bias $H_b$ and coercivity $H_c$.
Here $\rho(H_b, H_c)$ is the so-called ``Preisach function''.

An intuitive picture can be obtained by first considering symmetric
hysterons, having no bias, i.e. $H_b = 0$. Starting from a fully ``up''
polarized state and decreasing the field to a negative $H_R$ switches down
all symmetric hysterons with $H_c<|H_R|$. Reversing the direction of the
sweep and increasing the field from $H_R$ to $-H_R$ along a reversal curve
switches back every switched hysteron. Thus at $H=-H_R$ saturation is
reached, creating a kink in the magnetization.  Symmetric hysterons
therefore give rise to reversal-field memory.  However, this memory effect
will be detectable only if the number of symmetric hysterons is
macroscopic. This happens if $\rho(H_b, H_c)$ has a Dirac delta
singularity at $H_b=0$ and $H_c=|H_R|$. As the kink is observed in a range
of $H_R$ values, the singularities of the Preisach function form a
horizontal ridge along the $H_b=0$ axis for the corresponding range of
$H_c=|H_R|$ values.

Next we move beyond phenomenological approaches, but keep the insight
gained from the Preisach model.  We carry over the concept of symmetric
hysterons as ``symmetric clusters'' of the strongly interacting spins of the
EASG. A spin $S_i$ belongs to a symmetric cluster if $S_i$ flips down only
after all its neighbors have flipped down, and during the reverse sweep
$S_i$ flips up again only after all its neighbors have. Therefore, this
central spin $S_i$ experiences an effective local field which is symmetric
with respect to the change of direction of the external field, in analogy
to a symmetric hysteron.

Spins possessing {\it local spin-reversal symmetry} are candidates for
symmetric hysterons. By local spin-reversal symmetry we mean that the
local field $h_i$, felt by $S_i$ [Eq.~(\ref{eq:local_field})], is
perfectly reversed if the external field $H$ is reversed and all spins
coupled to $S_i$ are reversed as well. Every spin of the EASG has local
spin-reversal symmetry. However, in a glassy system the spin
configurations depend on the history of the sample.  Therefore, at $-H_R$
the neighbors of most spins {\it do not necessarily point in a direction
opposite of their direction at $H_R$}, and thus most EASG spins do not
belong to symmetric clusters. Hence the model Hamiltonian possessing a
local spin-reversal symmetry is a necessary but not sufficient condition
for having symmetric clusters.

To see a macroscopic kink it has to be shown that the density of symmetric
clusters is finite. A lower bound on their density is obtained by
considering the simplest symmetric cluster: two strongly coupled spins,
weakly coupled to their six neighbors in a 2D lattice.  The switching
field of each spin is determined by these couplings. The outer spins will
switch before the inner spins if their couplings are restricted by
appropriate inequalities, confining the couplings to finite intervals.  
The density of symmetric clusters is obtained by integrating the product
of the distributions of the couplings over these finite intervals. With
unbounded coupling distributions, e.g.~Gaussian, the product of the
distributions is finite over the finite integration intervals, thus the
resulting density is finite as well.

\begin{figure}
\epsfxsize=7.0cm
\begin{center}
\epsfbox{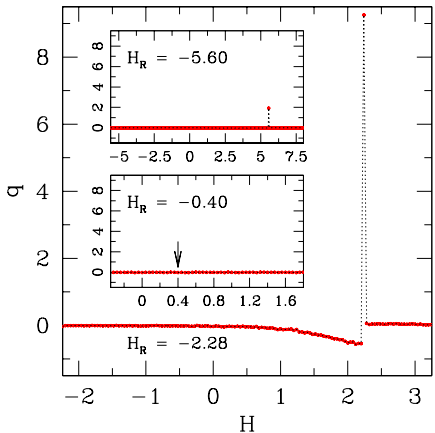}
\end{center}
\vspace*{-1.3cm}
\caption{\label{fig:overlap}
Overlap function $q$ of the spins flipping at $H_R$ and at $H > H_R$, for
$H_R = -2.28$. The insets show data for $H_R = -0.40$ and $H_R = -5.60$.
The arrow in the inset marks $H = 0.40$.}
\end{figure}

As a further evidence for the macroscopic number of symmetric clusters, we
define an overlap function $q$ between the spins which flip at $H_R$ and
the spins which flip at $H > H_R$:
\begin{eqnarray}
q(H) &=& \frac{1}{4}\sum_i[S_i(H_R + \delta) - S_i(H_R)] \times 
\nonumber \\
      &&\;\;\;\;\;\;\;\;\;\;\;\;\;\;\;\;\;[S_i(H + \delta) - S_i(H)] \; 
.
\label{overlap}
\end{eqnarray}
Here $\delta$ is the field step. In Fig.~\ref{fig:overlap} we show the
overlap $q(H)$ for $H_R = -2.28$. The large peak at $H = +2.28$ indicates
that a macroscopic number of spins which have flipped at $H_R$, also flip
at $-H_R$. This in turn means that there are a macroscopic number of
symmetric clusters. The insets show a much smaller number of symmetric
clusters at $H_R = -0.40$ and $H_R = -5.60$, values outside the peak of
Fig.~\ref{fig:delta}.

To characterize reversal-field memory further, we adapt a new tool
developed for analyzing experimental data of hysteretic
systems \cite{pike:99}. A family of First Order Reversal Curves (FORCs)
with different $H_R$ is generated, with $M(H, H_R)$ denoting the resulting
magnetization as a function of the applied and reversal fields.  
Computing the mixed second order derivative $\rho(H, H_R)= -(1/2)
[{\partial}^2 M/{\partial} H {\partial} H_R]$ and changing variables to
$H_c=(H-H_R)/2$ and $H_b=(H+H_R)/2$, the local coercivity and bias,
respectively, yields the ``FORC distribution'' $\rho(H_b, H_c)$. For
phenomenological Preisach models, the FORC distribution is equal to the
Preisach function.  However, FORC distributions are more general, because
they are extracted from numerical or experimental data, and thus are
model independent.

\begin{figure}
\epsfxsize=12cm
\begin{center}
\epsfbox{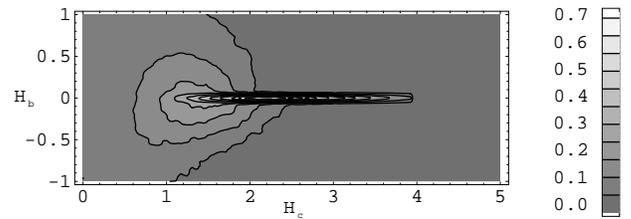}
\end{center}
\vspace*{-0.5cm}
\caption{\label{fig:forc-ea}
FORC Diagram of the EASG. Note the ridge along the $H_c$ axis.
}
\end{figure}

Figure \ref{fig:forc-ea} shows the FORC diagram of the EASG. The ridge
along the $H_c$ axis in the range $1.5 < H_{c} < 4.0$ corresponds to the
peak of Fig.~\ref{fig:delta}, representing the kinks of
Fig.~\ref{fig:kink}.  Thus FORC diagrams capture the reversal-field memory
effect in the form of a ridge along the $H_c$ axis.

\begin{figure}
\epsfxsize=1.6cm
\begin{center}
\epsfbox{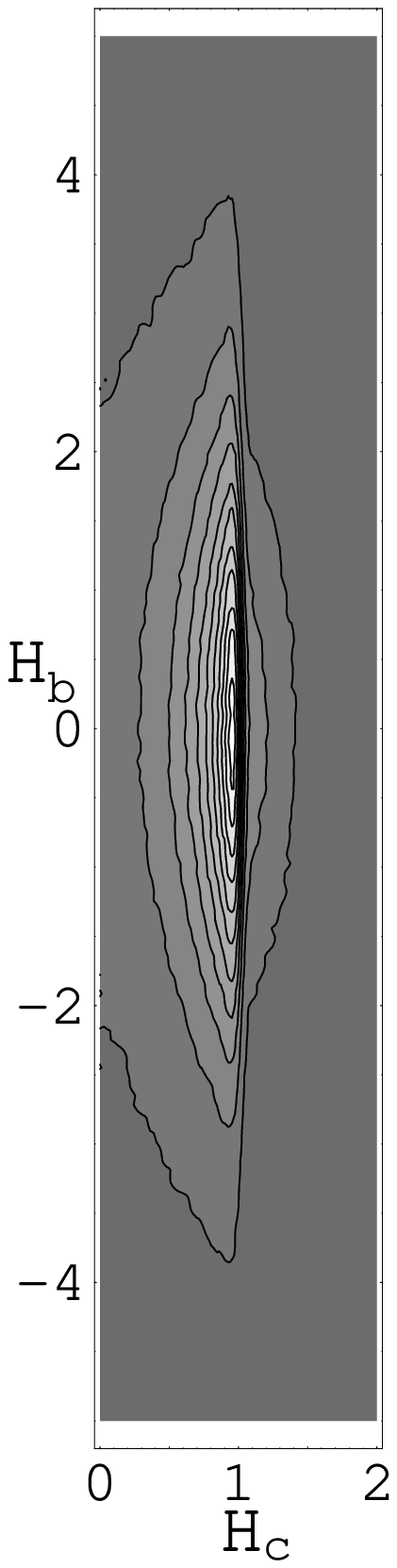}
\epsfxsize=7cm
\hspace*{-0.6cm}
\epsfbox{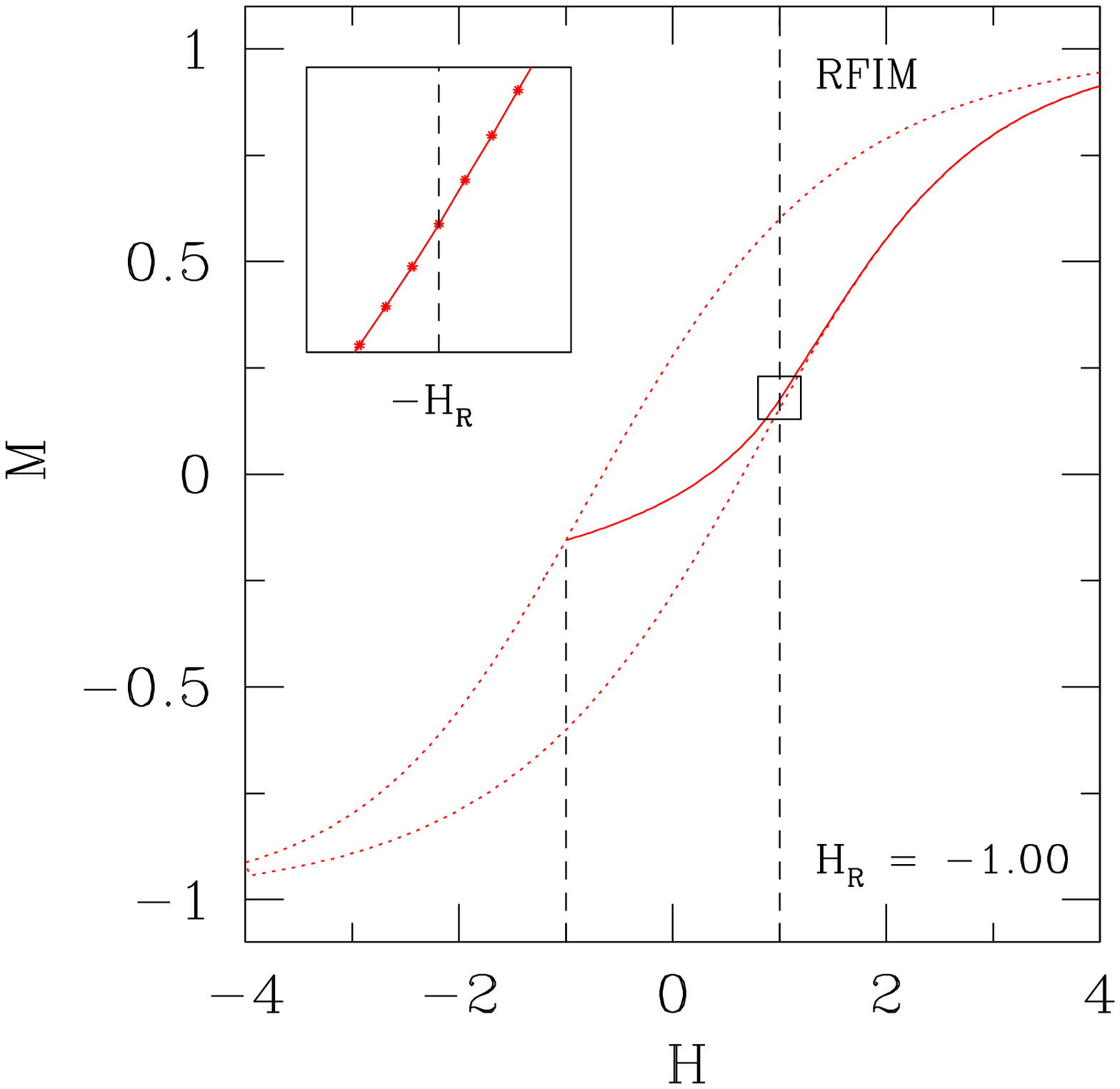}
\end{center}
\vspace*{-1.2cm}
\caption{\label{fig:forc-rfim}
FORC Diagram of the 2D RFIM. Data for $\Delta = 4.0$, $N =
10^4$ spins and $10^3$ disorder realizations (same scale as in
Fig.~/\ref{fig:forc-ea}). Note the clear differences to the FORC diagram
of the EASG (Fig.~\ref{fig:forc-ea}). The right panel
shows the major hysteresis loop (dotted line) and a reversal curve (solid
line) for $H_R = -1$. The inset shows an enlarged view of the region 
around $-H_R$.}
\end{figure}

To demonstrate that local spin reversal symmetry of the Hamiltonian is
necessary for reversal field memory to be present, we study the Random
Field Ising Model (RFIM) \cite{binder:86,sethna:93}. In this model $J_{ij}
= 1$ and the disorder is introduced through random local fields chosen
according to a Gaussian distribution with zero mean and standard deviation
$\Delta$. Direct inspection reveals that the RFIM {\it does not possess} a
local spin-reversal symmetry. Therefore, the RFIM cannot have symmetric
clusters and should not exhibit a reversal-field memory. This is confirmed
by our simulations: a typical RFIM reversal curve shown in the right
panel of Fig.~\ref{fig:forc-rfim} has no kink at $-H_R$.

Not only do the two models differ in the local spin symmetry, the EASG
possesses frustration which might give rise to hysteretic phenomena that
are qualitatively rather different than in the the RFIM.  To explore this
possibility, we show in the left panel of Fig.~\ref{fig:forc-rfim} the
FORC diagram for the RFIM for a disorder $\Delta = 4.0$.  While the major
hysteresis loop of the RFIM is very similar to that of the EASG, the FORC
distribution is qualitatively different: it exhibits a predominantly
vertical feature. The distribution of random fields of the RFIM introduces
a large range of biases for the spins, with little variation in the local
coercivity. This expectation is confirmed by simulations of the RFIM with
various disorder distributions, which show that the vertical cross section
of the vertical feature mirrors the shape of the random field distribution.

\begin{figure}
\epsfxsize=8cm
\begin{center}
\epsfbox{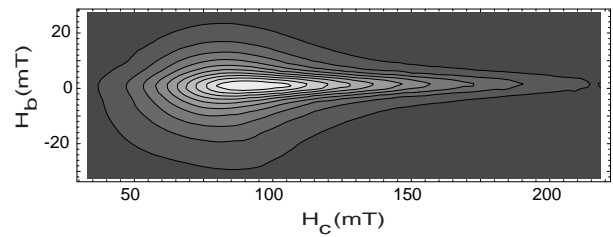}
\end{center}
\vspace*{-0.6cm}
\caption{\label{fig:kodak}
Experimental FORC diagram of a Kodak sample. Note the similarity to the
FORC diagram of the EASG shown in Fig.~\ref{fig:forc-ea}.}
\end{figure}

Finally we demonstrate the existence of reversal-field memory in
experimental systems. We study thin films of well-dispersed single-domain
magnetic Co-$\gamma$-Fe$_2$O$_3$ particles provided by Kodak Inc. We
determine both the individual reversal curves and the FORC diagram of the
system. While the reversal field memory kinks in $-H_R$ are somewhat
smoothed, the FORC diagram clearly exhibits the horizontal ridge
associated with the reversal-field memory effect (Fig.~\ref{fig:kodak}).
This striking similarity between the experimentally determined FORC
diagram of the Co-$\gamma$-Fe$_2$O$_3$ films and the numerically
determined FORC diagram of the EASG indicates not only that
Co-$\gamma$-Fe$_2$O$_3$ films exhibit reversal-field memory but also that
frustration may be a component of the physics of the
Co-$\gamma$-Fe$_2$O$_3$ films.

In conclusion, we have reported a novel reversal-field memory effect in
the EASG that manifests itself as a sharp kink in first order reversal
curves and also as a sharp ridge on the zero bias axis of FORC diagrams.
We suggest the microscopic origin of the effect is the presence of a
macroscopic number of ``symmetric clusters,'' and prove this by computing
a suitable overlap function. We further show that reversal field memory
is absent from the RFIM, which does not exhibit symmetric clusters. While
the hysteresis loops of the EASG and the RFIM are remarkably similar for
corresponding parameters, their FORC diagrams are profoundly different,
establishing that the FORC method is a powerful diagnostic tool for
capturing the sensitive details of hysteretic systems such as spin glasses
and random magnets. The FORC diagrams of several magnetic thin films
exhibit a profound ridge indicative of the reversal-field memory effect in
experimental systems. Simulations on more realistic magnetic models which
include dipolar interactions have also shown the reversal-field
memory effect. This suggests that the reversal-field
memory is not specific to the EASG, but it a robust result for a large
class of theoretical models and experimental systems. 

This work was supported by NSF Grant No.~DMR-9985978, No.~99-09468,
No.~EAR-99-09468, and No.~INT-9720440. 
We would like to thank T.~Jagielinski for 
characterization of the Kodak sample and D.~P.~Belanger and 
B.~A.~Allgood for discussions.

\bibliography{refs}

\begin{thebibliography}{9}
\expandafter\ifx\csname natexlab\endcsname\relax\def\natexlab#1{#1}\fi
\expandafter\ifx\csname bibnamefont\endcsname\relax
  \def\bibnamefont#1{#1}\fi
\expandafter\ifx\csname bibfnamefont\endcsname\relax
  \def\bibfnamefont#1{#1}\fi
\expandafter\ifx\csname citenamefont\endcsname\relax
  \def\citenamefont#1{#1}\fi
\expandafter\ifx\csname url\endcsname\relax
  \def\url#1{\texttt{#1}}\fi
\expandafter\ifx\csname urlprefix\endcsname\relax\def\urlprefix{URL }\fi
\providecommand{\bibinfo}[2]{#2}
\providecommand{\eprint}[2][]{\url{#2}}

\bibitem[{\citenamefont{Young}(1998)}]{young:98}
\bibinfo{editor}{\bibfnamefont{A.~P.} \bibnamefont{Young}}, ed.,
  \emph{\bibinfo{title}{Spin Glasses and Random Fields}}
  (\bibinfo{publisher}{World Scientific}, \bibinfo{address}{Singapore},
  \bibinfo{year}{1998}).

\bibitem[{\citenamefont{Bertotti}(1998)}]{bertotti:98}
\bibinfo{author}{\bibfnamefont{G.}~\bibnamefont{Bertotti}},
  \emph{\bibinfo{title}{Hysteresis and Magnetism for Physicists, Materials
  Scientists, and Engineers}} (\bibinfo{publisher}{Academic Press},
  \bibinfo{address}{New York}, \bibinfo{year}{1998}).

\bibitem[{\citenamefont{{Sethna} et~al.}(1993)\citenamefont{{Sethna}, {Dahmen},
  {Kartha}, {Krumhansl}, {Roberts}, and {Shore}}}]{sethna:93}
\bibinfo{author}{\bibfnamefont{J.~P.} \bibnamefont{{Sethna}}},
  \bibinfo{author}{\bibfnamefont{K.}~\bibnamefont{{Dahmen}}},
  \bibinfo{author}{\bibfnamefont{S.}~\bibnamefont{{Kartha}}},
  \bibinfo{author}{\bibfnamefont{J.~A.} \bibnamefont{{Krumhansl}}},
  \bibinfo{author}{\bibfnamefont{B.~W.} \bibnamefont{{Roberts}}},
  \bibnamefont{and} \bibinfo{author}{\bibfnamefont{J.~D.}
  \bibnamefont{{Shore}}}, \bibinfo{journal}{Phys. Rev. Lett.}
  \textbf{\bibinfo{volume}{70}}, \bibinfo{pages}{3347} (\bibinfo{year}{1993}).

\bibitem[{\citenamefont{{Lyuksyutov} et~al.}(1999)\citenamefont{{Lyuksyutov},
  {Nattermann}, and {Pokrovsky}}}]{lyuksyutov:99}
\bibinfo{author}{\bibfnamefont{I.~F.} \bibnamefont{{Lyuksyutov}}},
  \bibinfo{author}{\bibfnamefont{T.}~\bibnamefont{{Nattermann}}},
  \bibnamefont{and}
  \bibinfo{author}{\bibfnamefont{V.}~\bibnamefont{{Pokrovsky}}},
  \bibinfo{journal}{Phys. Rev. B} \textbf{\bibinfo{volume}{59}},
  \bibinfo{pages}{4260} (\bibinfo{year}{1999}).

\bibitem[{\citenamefont{Zhu}(1990)}]{zhu:90}
\bibinfo{author}{\bibfnamefont{J.}~\bibnamefont{Zhu}}, in
  \emph{\bibinfo{booktitle}{Magnetic Recording Technology}}, edited by
  \bibinfo{editor}{\bibfnamefont{C.~D.} \bibnamefont{Mee}} \bibnamefont{and}
  \bibinfo{editor}{\bibfnamefont{E.~D.} \bibnamefont{Daniel}}
  (\bibinfo{publisher}{Mc Graw Hill}, \bibinfo{address}{New York},
  \bibinfo{year}{1990}).

\bibitem[{\citenamefont{{P{\' a}zm{\' a}ndi} et~al.}(1999)\citenamefont{{P{\'
  a}zm{\' a}ndi}, {Zar{\' a}nd}, and {Zim{\' a}nyi}}}]{pazmandi:99}
\bibinfo{author}{\bibfnamefont{F.}~\bibnamefont{{P{\' a}zm{\' a}ndi}}},
  \bibinfo{author}{\bibfnamefont{G.}~\bibnamefont{{Zar{\' a}nd}}},
  \bibnamefont{and} \bibinfo{author}{\bibfnamefont{G.~T.} \bibnamefont{{Zim{\'
  a}nyi}}}, \bibinfo{journal}{Phys. Rev. Lett.} \textbf{\bibinfo{volume}{83}},
  \bibinfo{pages}{1034} (\bibinfo{year}{1999}).

\bibitem[{\citenamefont{Pike et~al.}(1999)\citenamefont{Pike, Roberts, and
  Verosub}}]{pike:99}
\bibinfo{author}{\bibfnamefont{C.~R.} \bibnamefont{Pike}},
  \bibinfo{author}{\bibfnamefont{A.~P.} \bibnamefont{Roberts}},
  \bibnamefont{and} \bibinfo{author}{\bibfnamefont{K.~L.}
  \bibnamefont{Verosub}}, \bibinfo{journal}{J. Appl. Phys.}
  \textbf{\bibinfo{volume}{85}}, \bibinfo{pages}{6660} (\bibinfo{year}{1999}).

\bibitem[{\citenamefont{Binder and Young}(1986)}]{binder:86}
\bibinfo{author}{\bibfnamefont{K.}~\bibnamefont{Binder}} \bibnamefont{and}
  \bibinfo{author}{\bibfnamefont{A.~P.} \bibnamefont{Young}},
  \bibinfo{journal}{Rev. Mod. Phys.} \textbf{\bibinfo{volume}{58}},
  \bibinfo{pages}{801} (\bibinfo{year}{1986}).

\bibitem[{\citenamefont{Preisach}(1935)}]{preisach:35}
\bibinfo{author}{\bibfnamefont{F.}~\bibnamefont{Preisach}},
  \bibinfo{journal}{Z. Phys.} \textbf{\bibinfo{volume}{94}},
  \bibinfo{pages}{277} (\bibinfo{year}{1935}).

\end{thebibliography}

\end{document}